\documentclass[trackchanges]{aastex63}
\usepackage{enumitem,kantlipsum}

\shorttitle{Subpulse Drifting in Partially Screened Gap}

\begin{document}

\title{Estimating the evolution of Sparks in Partially Screened Gap of Pulsars from Subpulse Drifting}


\author[0000-0003-1824-4487]{Rahul Basu}
\affiliation{Janusz Gil Institute of Astronomy, University of Zielona G\'ora, ul. Szafrana 2, 65-516 Zielona G\'ora, Poland.}

\author[0000-0002-9142-9835]{Dipanjan Mitra}
\affiliation{National Centre for Radio Astrophysics, Tata Institute of Fundamental Research, Pune 411007, India.}
\affiliation{Janusz Gil Institute of Astronomy, University of Zielona G\'ora, ul. Szafrana 2, 65-516 Zielona G\'ora, Poland.}

\author[0000-0003-1879-1659]{George I. Melikidze}
\affiliation{Janusz Gil Institute of Astronomy, University of Zielona G\'ora, ul. Szafrana 2, 65-516 Zielona G\'ora, Poland.}
\affiliation{Evgeni Kharadze Georgian National Astrophysical Observatory, 0301 Abastumani, Georgia.}

\begin{abstract}
A novel scheme has been developed to show that the observed phase behaviour 
associated with subpulse drifting from two pulsars, J1034$-$3224 and 
J1720$-$2933, can be used to obtain the magnetic field configuration in the 
partially screened gap (PSG). The outflowing plasma along the open magnetic 
field line region of pulsars is generated due to spark discharges in an inner 
acceleration region (IAR) above the polar cap. The IAR has been modelled as a 
partially screened gap (PSG) with a steady supply of positively charged ions 
emitted from the heated polar cap surface dominated by strong non-dipolar 
magnetic fields. In a PSG the sparks are tightly packed and constrained to be 
present along the polar cap boundary. The sparks lag behind the rotation of the
star during their lifetimes. As a result the sparking pattern evolves along two
different directions in the clockwise and counter-clockwise manner around a 
stationary central spark, and can be associated with the observed phenomenon of
subpulse drifting. PSR J1034$-$3224 has four prominent components and exhibit 
bi-drifting where alternate components show opposite sense of drifting, while 
PSR J1720$-$2933 has a single component profile and shows systematic 
coherent drift bands. We show that the differences in their drifting behaviour
can be directly linked to different natures of the non-dipolar surface magnetic
field configurations. 
\end{abstract}


\section{Introduction}

A steady outflow of relativistic plasma is setup along the open magnetic field
lines of the pulsar magnetosphere. This outflowing plasma forms the pulsar wind
and is also the source of coherent radio emission which arises due to nonlinear
plasma instabilities. Detailed observations have shown that the radio emission 
is emitted close to the surface at heights less than 10 percent of the light 
cylinder radius \citep{vHX97,KG98,ET_MR02,WJ08,KMG09,M17}. This requires the 
outflowing plasma to be generated primarily in an inner acceleration region 
(IAR) above the polar caps. The IAR was initially modelled as a inner vacuum 
gap (IVG), where the gap extended to a height of around 100 meters above the 
surface \citep{RS75}. The plasma is generated from spark discharges due to 
cascading electron-positron pair production in the gap, which are subsequently 
accelerated to relativistic energies in the large potential drop, with the
positrons flowing outward and the electrons accelerated backward to heat the 
surface. The cascading effect requires the surface magnetic field to be highly 
non-dipolar in nature, while heating from the backstreaming particles raises 
the surface temperatures above the polar cap to around $10^6$ kelvins. The high
surface temperatures and presence of non-dipolar fields on the polar cap have 
been confirmed by observations of X-ray emission from the stellar surface 
\citep{GHM08,H13,SGZ17,G17,HKB18,AM19,SG20,PM20}. The surface temperatures are 
close to the critical level such that the high energy tail of the distribution 
of ions can flow freely from the surface and form a partially screened gap 
\citep[PSG, see section 2.1][]{GMG03}. In a PSG, spark discharges are a 
mechanism of regulating the surface temperature around this critical level and 
the sparks are constrained to be arranged around the polar cap boundary. The 
sparks are formed in a tightly packed configuration across the cross section of
the polar cap and lag behind the rotation of the pulsar due to 
\textit{\textbf{E}}\textbf{x}\textit{\textbf{B}} drift in the gap \citep{MBM20,
BMM20b,BMM22}.

The phenomenon of subpulse drifting, seen in the single pulse sequence of 
several pulsars, reflects the drift behaviour of the sparks \citep{WES06,
BMM16,SWS23}, and provides a direct probe into the physical processes within 
the IAR. The drifting behaviour is measured using the longitude resolved 
fluctuation spectrum (LRFS) analysis, where Fourier transforms of the pulse 
sequence are carried out along several narrow longitude ranges within the 
emission window to characterise the periodic behaviour \citep{B73}. The primary
measurements are the drifting periodicity, $P_3$, that can be related to the 
electric potential difference in the IAR, and the phase variations across the 
emission window, which follows the evolution of the sparking pattern in the 
non-dipolar polar cap along the observers' line of sight (LOS). The drifting 
phase behaviour shows different patterns ranging from systematic change across 
the profile to phase shifts in different components \citep{BMM19}. The most 
remarkable behaviour is seen in a small category of pulsars that exhibit 
bi-drifting, where the drifting phases show reversal in directions in different
components of the profile \citep{CLM05,W16,SvL17,BM18a,BPM19,SvLW20,SBD22}.

Modelling the observed drifting phase behaviour requires estimating the surface
magnetic field configuration that determines the polar cap structure, and the 
temporal evolution of the sparking pattern in the IAR above the polar cap. 
\citet{BMM22} showed that in a PSG the sparking pattern is constrained by the 
boundary of the polar cap to evolve around concentric rings in a direction 
opposite to the rotation of the pulsar. Although the individual sparks have 
much shorter durations than the drifting periodicity, the subsequent sparks are
formed shifted either in the clockwise or counter-clockwise manner around a 
stationary spark at the center and the overall sparking pattern mimics a drift 
behaviour. The non-dipolar surface magnetic field configuration on the other 
hand is less well understood. \citet{GMM02} suggested a simple approximation 
consisting of a star centered dipole and a weaker local dipole on the surface 
that determines the surface field. 

In this work we have used the measured drifting behaviour from two pulsars 
J1034$-$3224 and J1720$-$2933 \citep{BM18a} with different characteristics, to 
estimate the spark behaviour in the PSG and the surface magnetic field 
configuration. PSR J1034$-$3224 has four prominent components in the average 
profile and shows the bi-drifting behaviour with alternate components having 
opposite drift directions. PSR J1720$-$2933 has a single component profile and 
show systematic drifting with large monotonic phase variations. These studies 
provide observational validation of the PSG model and give an outline for 
estimating the physical properties of IAR in pulsars. In section \ref{sec:obs} 
we report the estimates of line of sight geometry from the observations in both
pulsars while section \ref{sec:psg} details the estimation of spark evolution 
in PSG and the nature of surface magnetic fields in order to reproduce the 
observed drifting behaviour. We summarise our results and briefly discuss the 
application of these studies in the larger pulsar population in section 
\ref{sec:con}.

\section{Radio Emission properties}\label{sec:obs}
\noindent
PSR J1034$-$3224 and PSR J1720$-$2933 were observed as part of the 
Meterwavelength Single Pulse Emission Survey \citep[MSPES,][]{MBM16} conducted 
using the Giant Meterwave Radio Telescope. Around 2000 polarized single pulses 
were observed in this survey at two separate frequencies of 325 MHz and 610 
MHz. Subsequently, longer observations were carried out by \citet{BM18a} at 325
MHz to characterize the drifting features in each pulsar. The drifting 
behaviour of the sparks is imprinted on the outflowing plasma which gives rise 
to the radio emission at heights of several hundred kilometers from the 
surface, where the magnetic field is dipolar in nature. Hence, it is imperative
to understand the dipolar geometry of the radio emission region to unravel the 
spark evolution in the IAR.

The quantities used to specify the geometry include the inclination angle 
between the rotation and magnetic axis, $\alpha$, angle at the closest approach
of the line of sight (LOS) to the magnetic axis, $\beta$, the emission height 
at a given frequency ($\nu$), $h_{\nu}$, the beam opening angle at that 
frequency, $\rho_{\nu}$, and the relative LOS traverse across the emission 
beam, $S_{los}=\beta/\rho_{\nu}$. The inclination angle can be obtained from 
the width of the profile components as \citep{SBM18} :
\begin{equation}
W_C = W_{B} P^{-0.5}/\sin{\alpha}
\label{eq:alpha}
\end{equation}
Here $W_C$ is the half width of the component in the profile at the observing
frequency. The distribution of the component widths as a function of period 
($P$), at any given frequency, are seen to be above a lower boundary line which
is specified by width parameter, $W_{B}$. The above estimate has a degeneracy 
between $\alpha$ and $\pi-\alpha$. The angles $\alpha$ and $\beta$ can be 
estimated from the rotating vector model \citep[RVM,][]{RC69} fits to 
polarization position angle (PPA). However, these yield correlated values which
are unreliable and only the steepest gradient point ($R_{ppa}$) of the PPA can 
be used to constrain the geometry as \citep{ML04} :
\begin{equation}
R_{ppa} = \vert\sin{\alpha}/\sin{\beta}\vert. 
\end{equation}
The beam opening angle can be estimated from the full width of the pulsar
profile, $W_{5\sigma}$, which is assumed to be connected with the last open 
field lines, using spherical geometry \citep{GGR84} :
\begin{equation}
\sin^2{(\rho_{\nu}/2)} = \sin{\alpha}\sin{(\alpha+\beta)}\sin^2{(W_{5\sigma}/4)} + \sin^2{(\beta/2)}
\label{eq:rho}
\end{equation}
The beam opening angle can be further used to estimate the radio emission 
height 
\begin{equation}
h_{\nu} = 10 P \left(\frac{\rho_{\nu}}{1.23\degr}\right)^2 {\rm km,}
\label{eq:emht}
\end{equation}
where the beam opening angle at distance of 10 km for the dipolar field lines
is 1.23\degr, for $P=1$ s.

\begin{deluxetable}{cccccccccccccc}
\tablecaption{Radio Emission Properties and Line of Sight Geometry \label{tab:geom}}
\tablewidth{0pt}
\tablehead{
  \colhead{PSR}  & \colhead{$P$} & \colhead{$\dot{P}$} & \colhead{$\nu$} & \colhead{$W_{C}$} & \colhead{$W_{5\sigma}$} & \colhead{$W_B$} & \colhead{$R_{ppa}$} & \colhead{$\alpha$} & \colhead{$\alpha_m$} & \colhead{$\beta$} & \colhead{$\rho$} &\colhead{$S_{los}$} & \colhead{$h$} \\
   & \colhead{(s)} & \colhead{($s~s^{-1}$)} & \colhead{(MHz)} & \colhead{(\degr)} & \colhead{(\degr)} & \colhead{(\degr)} & \colhead{($\degr$)} & \colhead{($\degr$)} & \colhead{($\degr$)} & \colhead{(\degr)} & \colhead{(\degr)} &   & (km)}
\startdata
 1034$-$3224 & 1.15 & $2.3\times10^{-16}$ & 325 & 7.4$\pm$0.9 & 80.2$\pm$1.8 & 2.37 & 9.95 & 17.4$\pm$2.0 & 16.6/163.4 & $\pm1.6$ & 11.9 & $\pm0.14$ & 1073 \\
   &   &   & 610 & 7.1$\pm$0.2 & 68.9$\pm$0.4 & 2.16 &   & 16.5$\pm$0.5 &   &   & 10.3 & $\pm0.16$ & 806 \\
   &   &   &   &   &   &   &   &   &   &   &   &   &  \\
 J1720$-$2933 & 0.62 & $7.5\times10^{-16}$ & 325 & 5.0$\pm$0.2 & 25.7$\pm$0.4 & 2.37 & -6.6 & 37.1$\pm$1.7 & 38.3/141.7 & $\pm5.4$ & 9.2 & $\pm0.59$ & 348 \\
   &   &   & 610 & 4.2$\pm$0.2 & 24.1$\pm$0.4 & 2.16 &   & 40.3$\pm$2.3 &   &   & 8.8 & $\pm0.61$ & 320 \\
\enddata
\end{deluxetable}

Table \ref{tab:geom} shows the estimates of geometry and emission heights at 
325 MHz and 610 MHz for the two pulsars J1034$-$3224 and J1720$-$2933. The 
profile and component widths at both frequencies were estimated in 
\citet{SBM18}, which also reported the estimate of $W_B$ from distribution of 
the widths. The PPA were reported in \citet{MBM16}, and we carried out RVM fits
to obtain $R_{ppa}$ for each pulsar (the detailed RVM fitting process for the 
pulsars in the MSPES survey, including the two reported here, is shown in Mitra
et al. 2023, in preparation). The $\alpha$ values obtained independently at the
two frequencies are consistent within measurement errors and a weighted mean 
$\alpha_m$ is used for estimating the other parameters.

\section{Estimating Properties of Partially Screened Gap from Subpulse Drifting} \label{sec:psg}
The surface magnetic field configuration is approximated by considering the 
presence of a dominant star centered dipole and one or more weaker dipoles 
located just below the stellar surface \citep{GMM02}, such that the magnetic 
field becomes dipolar a few kilometers away from the surface. The polar 
cap configuration is determined by the relative strength and location of the
surface dipoles. The presence of subpulse drifting suggests the polar cap to 
have a relatively smooth boundary and can be approximated to have an elliptical
shape with major axis $a_{cap}$, minor axis $b_{cap}$ and inclination angle 
$\theta_{cap}$ in a given co-ordinate plane. The center of the polar cap on the
neutron star surface is located at $(R_S, \theta_{cap}^c, \phi_{cap}^c)$, in
the rotating co-ordinate system with the neutron star at the center and the
rotation axis aligned along the z-axis. Here, $R_S=10^6$ cm is the radius of 
the neutron star. For a given polar cap, the spark sizes are constrained from 
average emission beam studies which have been shown to comprise of a central 
core component surrounded by two concentric rings of conal emission 
\citep[][also see appendix \ref{app:psgsprk}]{ET_R93,MD99}. The potential drop 
across the sparks is expected to vary along the different axes in elliptical 
polar caps and the spark shape resemble the polar cap, specified by major axis 
$a_{spark}$, minor axis $b_{spark}$, with effective size $h_{\perp}\sim
\sqrt{a_{sprk} b_{sprk}}$. A detailed characterization of the PSG can be 
achieved from the drifting periodicity $P_3$, and the non-dipolar surface 
fields in the polar cap characterized by the parameters $b=B_s/B_d$, $B_s$ 
being the non-dipolar field strength and $B_d$ the global dipolar field, and 
$\alpha_l$, the angle made by the local non-dipolar magnetic field with the 
rotation axis \citep{MBM20,BMM22}. The screening factor $\eta$ of the potential
drop in a PSG is obtained as 
\begin{equation}
\eta = 1/(2\pi P_3 |\cos{\alpha_l}|),
\end{equation}
The electric potential difference in the gap can be estimated as 
\begin{equation}
\Delta V_{PSG} = \frac{4 \pi \eta b B_d |\cos{\alpha_l}|}{P c} h_{\perp}^2
\end{equation}
and the surface temperature of the polar cap, which corresponds to the critical
temperature $T_i$, is 
\begin{equation}
T_i = (\eta b)^{1/2}~|\cos{\alpha_l}|^{1/4}~\left(\frac{h_{\perp}}{2.6 {\rm m}}\right)^{1/2}~\left(\frac{\dot{P}_{-15}}{P}\right)^{1/4}  \times 10^6 ~~ {\rm K}.
\end{equation}
Here $\dot{P}_{-15}$ the period derivative in units of $10^{-15} s s^{-1}$. A 
detailed scheme of estimating the sparking pattern evolution in PSG and the 
corresponding subpulse drifting behaviour for a specific surface field 
configuration has been discussed in \citet{BMM20b,BMM22}. We would like to 
emphasize that in this work we are concerned with inverting the problem to 
obtain constraints for the surface magnetic field from the observed drifting 
behaviour. 

The magnetic field configuration is modelled using a combination of a global 
dipole located at the origin and one or more local crust-anchored dipoles 
\citep{GMM02}. The global dipole is oriented in the x-z plane and specified as 
\textit{\textbf{d}} = ($d, \theta_d, 0\degr$), here $d=B_d R_S^3$, 
$B_d=10^{12}(P\dot{P}_{-15})^{0.5}$ G, and $\theta_d=\alpha$, obtained from 
eq.(\ref{eq:alpha}). The crust-anchored dipoles which are less well 
constrained, with dipole moments \textit{\textbf{m$_i$}} = ($m^i, \theta_m^i, 
\phi_m^i$), where i = 1, 2, ..., located at \textit{\textbf{r$_i$}} = ($r_s^i, 
\theta_s^i, \phi_s^i$) just below the surface. Although, it is likely that a 
number of effective dipoles exist near the surface, the non-dipolar polar cap 
is expected to be formed around one dominant dipole. Observations of X-ray 
emission from the surface of a number of pulsars show the magnetic field 
strength to be in the range $b\sim10-100$ \citep{G17,SG20}. The magnetic field 
lines from the non-dipolar polar cap surface connects with the dipolar field in
the emission region at a height $h_{\nu}\sim100-1000$ km from the surface. 
Depending on the relative orientation of the polar cap, the LOS gets twisted 
and can have different cuts on the surface for the same emission geometry, 
specified by $\alpha$ and $\beta$. The phase variations associated with 
drifting, which measures the delay between the spark peaks along the LOS cut of
the polar cap surface, exhibit a unique behaviour for every different 
orientation of the surface dipole. 

We have carried out a numerical search for the surface magnetic field 
configuration of the two pulsars J1034$-$3224 and J1720$-$2933. The parameters
of \textit{\textbf{m}} and \textit{\textbf{r}} were varied to obtain different 
realizations of the polar cap. The drifting phase behaviour for the LOS 
geometry in each polar cap was estimated and subsequently compared with the 
measured drifting behaviour to find unique solutions. We fixed $m=0.01d$ and 
$r_s=0.95R_S$ to ensure the appropriate range of $b$ and also reduce the 
parameter space for the search. The results of the polar cap fits and the 
corresponding PSG properties in each pulsar is presented below.

\begin{deluxetable}{cccccccccccc}
\tablecaption{The physical parameters of Partially Screened Gap \label{tab:capfit}}
\tablewidth{0pt}
\tablehead{
    & \colhead{$a_{cap}$} & \colhead{$b_{cap}$} & \colhead{$\theta_{cap}$} & \colhead{$\theta_{cap}^c$} & \colhead{$\phi_{cap}^c$} & \colhead{$b=B_s/B_d$} & \colhead{$|\cos{\alpha_l}|$} & \colhead{$\eta$} & \colhead{$h_{\perp}$} & \colhead{$T_i$} & \colhead{$\Delta V_{PSG}$} \\
   & \colhead{(m)}  & \colhead{(m)} & \colhead{($\degr$)} &  \colhead{($\degr$)} & \colhead{($\degr$)} &   &   &   & \colhead{(m)} & \colhead{($10^6$ K)} & \colhead{($10^{10}$ V)}}
\startdata
 J1034$-$3224 & 36.2 & 15.1 & -45.5 & 166.5 & -48.8 & $\sim67$ & $\sim0.65$ & 0.034 & 4.3 & 1.17 & 1.56 \\
   &   &   &   &   &   &   &   &   &   &   &   \\
 J1720$-$2933 & 75.2 & 30.1 & -36.8 & 37.1 & 20.9 & $\sim32$ & $\sim0.25$ & 0.26 & 8.8 & 3.93 & 22.4 \\
   &   &   &   &   &   &   &   &   &   &   &   \\
\enddata
\end{deluxetable}

\begin{figure}
\gridline{\fig{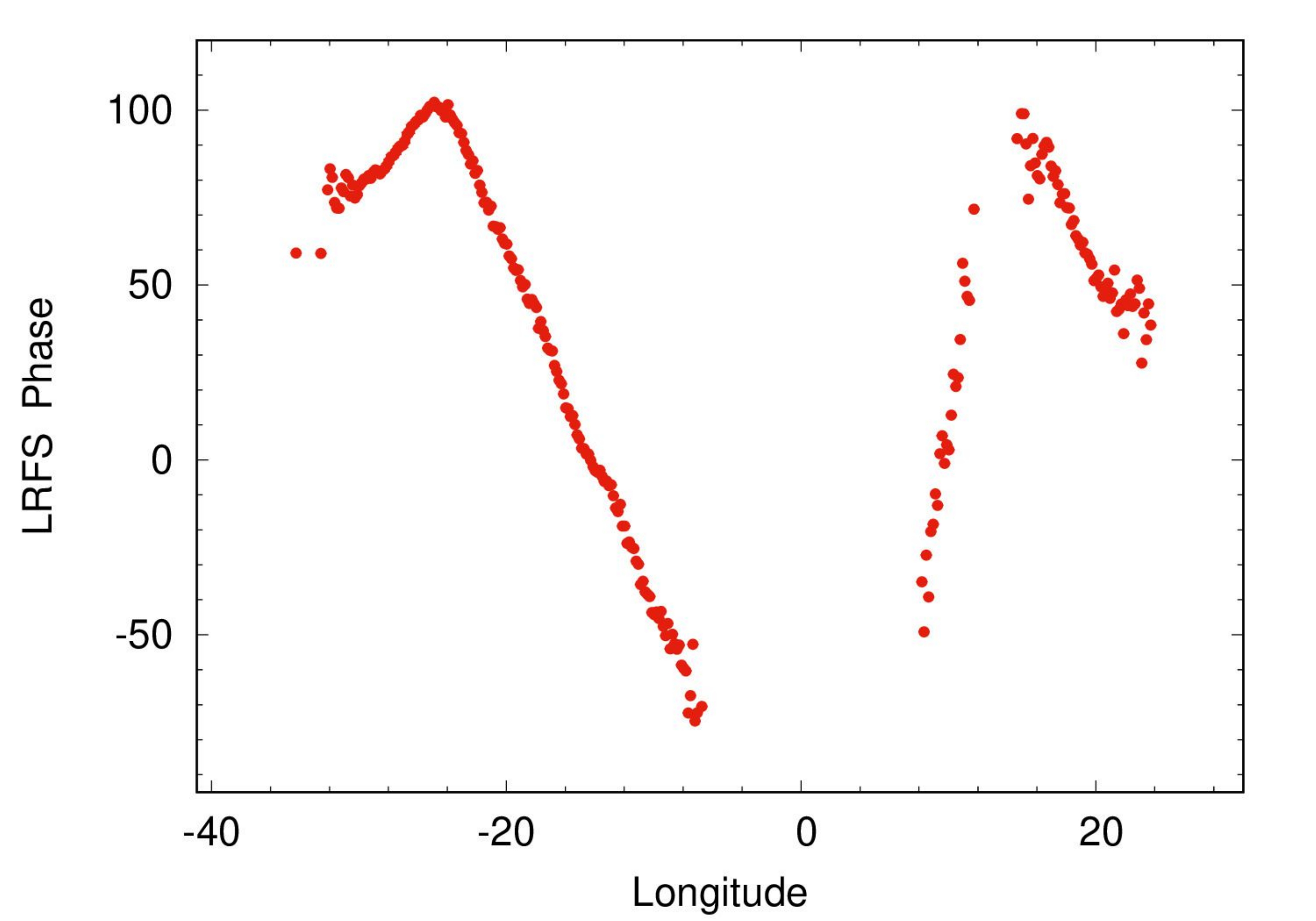}{0.45\textwidth}{(a)}
          \fig{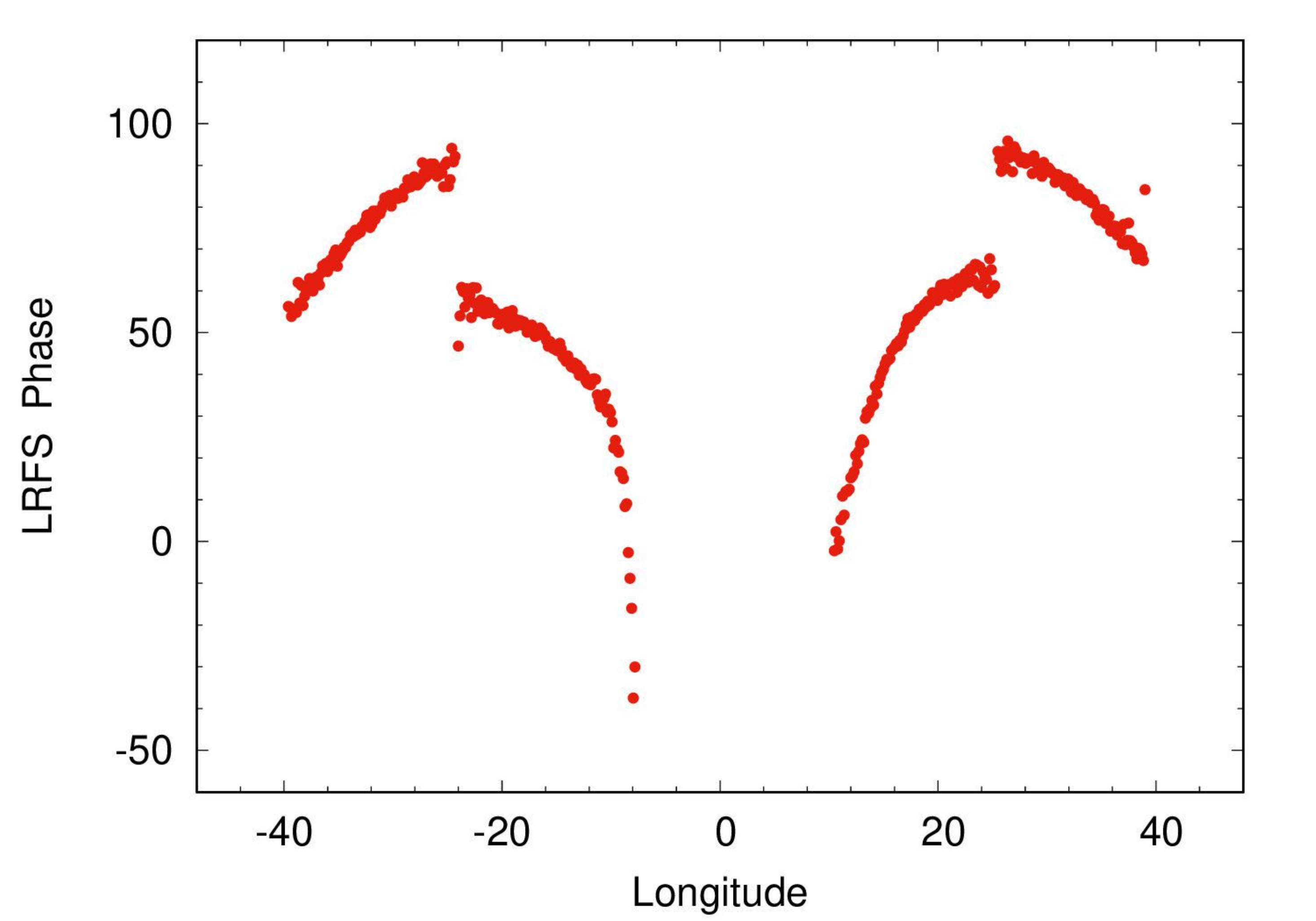}{0.45\textwidth}{(b)}}
\caption{(a) The observed phase behaviour associated with subpulse drifting in 
PSR J1034$-$3224, reported in \citet{BM18a}. The pulsar has four profile 
components, and the subpulse drifting shows the rare bi-drifting nature in this
pulsar. The first and third components have a positive slopes in the phase 
variations, while the second and fourth components have negative slopes. The 
two inner components have steeper slopes compared to the two outer ones. The 
phases in the first two components have been shifted by +180$\degr$~while the 
third component has been shifted by -180$\degr$~to provide optimum range along 
the y-axis for highlighting the bi-drifting behaviour in this pulsar. (b) The 
estimated phase behaviour associated with subpulse drifting in a simulated 
single pulse sequence is shown. The phase variations reproduce the primary 
features seen in PSR J1034$-$3224, including the bi-drifting nature and the 
relative slope changes between the inner and outer components.
\label{fig:LRFSphs_J1034}}
\end{figure}

\paragraph{PSR J1034$-$3224} The pulsar has four primary components in the 
profile as well as a pre-cursor which becomes prominent at higher frequencies 
above 600 MHz \citep{BMR15}. The observed radio emission properties and the
estimated LOS geometry is shown in Table \ref{tab:geom}. Subpulse drifting is
seen in the four components of the profile with drifting periodicity 
$P_3=7.2\pm0.7 P$ and the phase variations exhibit bi-drifting behaviour	
\citep[see Fig\ref{fig:LRFSphs_J1034}a,][]{BM18a}. The first and third 
components show positive slope in their phase variations while the second and 
fourth components have negative slopes. The phase variations are steeper
by roughly a factor of two in the inner components compared with the outer 
ones. We obtained the best fit surface magnetic field configuration to 
reproduce the bi-drifting behaviour in this pulsar to be \textit{\textbf{m}} = 
($0.01d, 165.9\degr, 312.0\degr$), located at \textit{\textbf{r}} = ($0.95R_S, 
133.5\degr, 237\degr$) for the emission geometry specified by 
$\alpha=163.4\degr$ and $\beta=1.6\degr$. The details of the polar cap 
parameters for this configuration as well as the spark distribution in the PSG 
is reported in Table \ref{tab:capfit} and appendix \ref{app:psgsprk}. A 
sequence of 200 single pulses were simulated for this configuration 
\citep[see][and appendix \ref{app:sinlrfs}]{BMM22} and the LRFS was estimated 
for this sequence. The phase behaviour of the LRFS is shown in 
Fig.\ref{fig:LRFSphs_J1034}b and reproduces the primary features of the 
bi-drifting behaviour seen in this pulsar.

\begin{figure}
\gridline{\fig{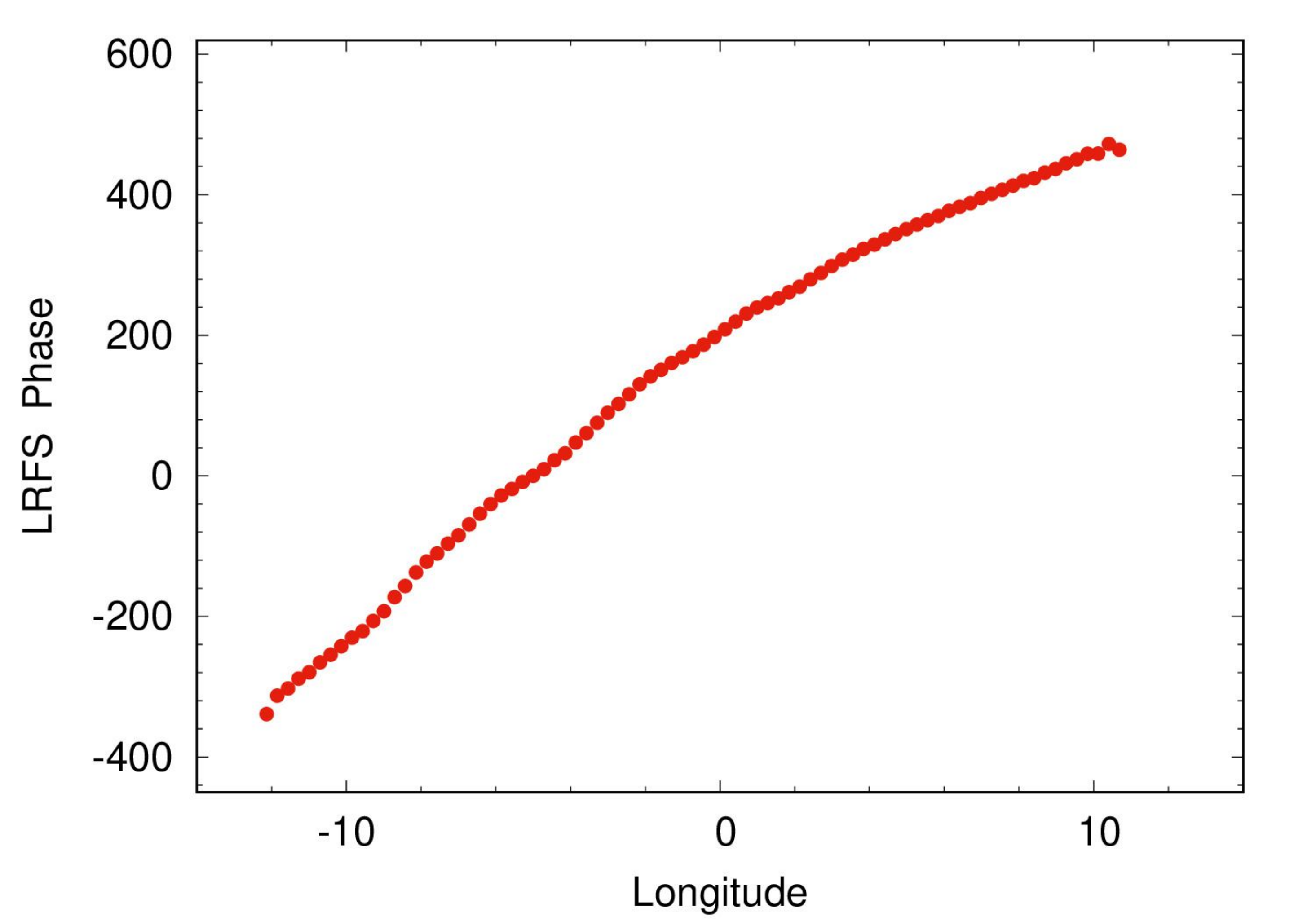}{0.45\textwidth}{(a)}
          \fig{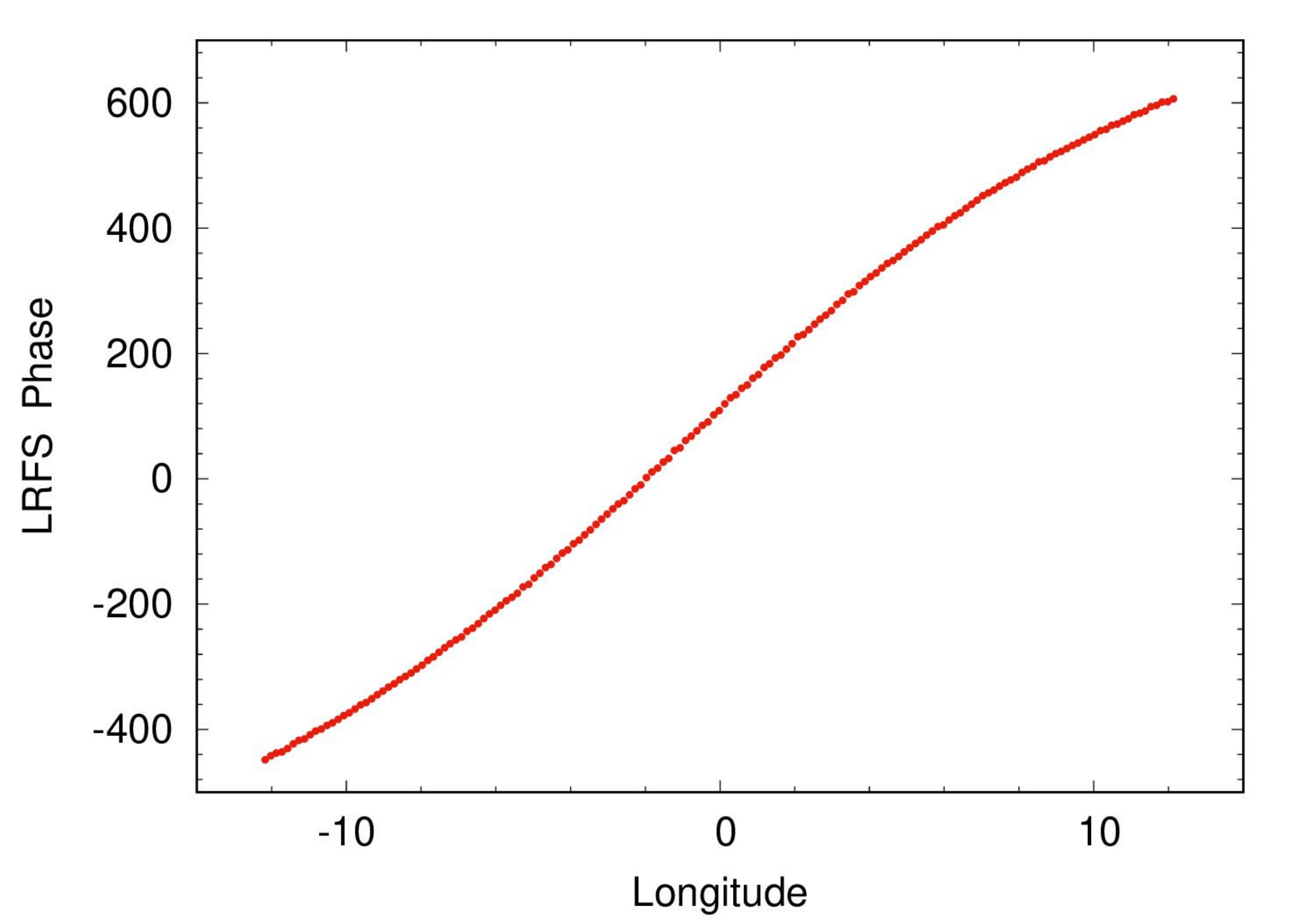}{0.45\textwidth}{(b)}}
\caption{(a) The observed phase behaviour associated with subpulse drifting in
PSR J1720$-$2933, reported in \citet{BM18a}. The phases show large variations 
across the profile with a positive slope across the phase window. (b) The 
estimated phase behaviour associated with subpulse drifting in a simulated 
single pulse sequence is shown. The phase variations exhibit the general 
observed behaviour seen in PSR J1720$-$2933.
\label{fig:LRFSphs_J1720}}
\end{figure}

\paragraph{PSR J1720$-$2933} The pulsar has one broad component in the profile
with a dip at the center and shows systematic coherent drifting across the
emission window. The drifting periodicity is $P_3=2.45P$ and the phases show 
large monotonic variations and slight flattening towards the trailing edge 
\citep[see Fig\ref{fig:LRFSphs_J1720}a,][]{BM18a}. The observed radio emission 
properties and the estimated LOS geometry is shown in Table \ref{tab:geom}. We 
obtained the best fit surface magnetic field configuration to be 
\textit{\textbf{m}} = ($0.01d, 36.3\degr, 24.0\degr$), located at 
\textit{\textbf{r}} = ($0.95R_S, 52.0\degr, 45.0\degr$) for the emission 
geometry specified by $\alpha=38.3\degr$ and $\beta=-5.4\degr$. The details of 
the polar cap parameters for this configuration as well as the spark 
distribution in the PSG is reported in Table \ref{tab:capfit} and appendix 
\ref{app:psgsprk}. We estimated the LRFS for a sequence of 200 single pulses 
that were simulated for the above configuration \citep[see][and appendix 
\ref{app:sinlrfs}]{BMM22}. The phase behaviour of the LRFS is shown in 
Fig.\ref{fig:LRFSphs_J1720}b and reproduces the general coherent drifting 
behaviour of this pulsar.

\section{Summary \& Conclusion} \label{sec:con}
\noindent
We have used the subpulse drifting behaviour from two contrasting examples of 
PSR J1034$-$3224 and PSR J1720$-$2933 to demonstrate that the evolution of the
spark distribution predicted by the PSG model of the IAR can be used to 
understand the nature of this difference. There are two primary factors that 
determine phase variations associated with subpulse drifting, the LOS geometry 
and the orientation of the polar cap over which the sparking pattern evolves. 
The LOS geometry determines the number of components in the average profile, 
with single and double component profiles associated with outer LOS cuts across
the emission beam with $|S_{los}| > 0.5$, while four or five component profiles
are seen in central cuts with $|S_{los}| < 0.2$. The systematic coherent drift
is usually seen in pulsars with single and double component profiles, while 
phase shifts and reversals are associated with profiles having both inner and 
outer conal components \citep{BMM19}. The LOS cut across the emission region
is altered during transition to the non-dipolar polar cap. The exact phase 
behaviour is determined by the effective LOS traverse across the polar cap and 
traces its relative orientation. We have used the drifting phase variations in 
the two pulsars to constrain the nature of the polar cap and the surface 
magnetic field configuration of the two pulsars. This allowed us to estimate 
the physical properties of the PSG and the spark distribution. Due to multiple 
parameters involved in specifying the surface dipole configuration the 
solutions reported here are not unique and there are other possible setups 
which can give similar drifting phase behaviour. But many of these solutions 
lead to unphysical results, like the polar cap being highly extended along an 
axis or having much higher $b$ value and located much further from the dipolar 
polar cap, and can be discarded as likely solutions. The X-ray observations of 
the stellar surface coupled with the drifting phase behaviour would enable 
tighter constraints on these estimations of the surface magnetic fields in 
neutron stars.


\section*{Acknowledgments}
DM acknowledges the support of the Department of Atomic Energy, Government of 
India, under project no. 12-R\&D-TFR-5.02-0700. DM acknowledges funding from 
the grant ``Indo-French Centre for the Promotion of Advanced Research - 
CEFIPRA" grant IFC/F5904-B/2018. This work was supported by the grant 
2020/37/B/ST9/02215 of the National Science Centre, Poland.

\bibliography{reflist}{}

\begin{thebibliography}{}
\expandafter\ifx\csname natexlab\endcsname\relax\def\natexlab#1{#1}\fi
\providecommand{\url}[1]{\href{#1}{#1}}
\providecommand{\dodoi}[1]{doi:~\href{http://doi.org/#1}{\nolinkurl{#1}}}
\providecommand{\doeprint}[1]{\href{http://ascl.net/#1}{\nolinkurl{http://ascl.net/#1}}}
\providecommand{\doarXiv}[1]{\href{https://arxiv.org/abs/#1}{\nolinkurl{https://arxiv.org/abs/#1}}}

\bibitem[{{Arumugasamy} \& {Mitra}(2019)}]{AM19}
{Arumugasamy}, P., \& {Mitra}, D. 2019, \mnras, 489, 4589,
  \dodoi{10.1093/mnras/stz2299}

\bibitem[{{Backer}(1973)}]{B73}
{Backer}, D.~C. 1973, \apj, 182, 245, \dodoi{10.1086/152134}

\bibitem[{{Basu} {et~al.}(2022){Basu}, {Melikidze}, \& {Mitra}}]{BMM22}
{Basu}, R., {Melikidze}, G.~I., \& {Mitra}, D. 2022, \apj, 936, 35,
  \dodoi{10.3847/1538-4357/ac8479}

\bibitem[{{Basu} \& {Mitra}(2018)}]{BM18a}
{Basu}, R., \& {Mitra}, D. 2018, \mnras, 475, 5098,
  \dodoi{10.1093/mnras/sty178}

\bibitem[{{Basu} {et~al.}(2020){Basu}, {Mitra}, \& {Melikidze}}]{BMM20b}
{Basu}, R., {Mitra}, D., \& {Melikidze}, G.~I. 2020, \mnras, 496, 465,
  \dodoi{10.1093/mnras/staa1574}

\bibitem[{{Basu} {et~al.}(2016){Basu}, {Mitra}, {Melikidze}, {Maciesiak},
  {Skrzypczak}, \& {Szary}}]{BMM16}
{Basu}, R., {Mitra}, D., {Melikidze}, G.~I., {et~al.} 2016, \apj, 833, 29,
  \dodoi{10.3847/1538-4357/833/1/29}

\bibitem[{{Basu} {et~al.}(2019{\natexlab{a}}){Basu}, {Mitra}, {Melikidze}, \&
  {Skrzypczak}}]{BMM19}
{Basu}, R., {Mitra}, D., {Melikidze}, G.~I., \& {Skrzypczak}, A.
  2019{\natexlab{a}}, \mnras, 482, 3757, \dodoi{10.1093/mnras/sty2846}

\bibitem[{{Basu} {et~al.}(2015){Basu}, {Mitra}, \& {Rankin}}]{BMR15}
{Basu}, R., {Mitra}, D., \& {Rankin}, J.~M. 2015, \apj, 798, 105,
  \dodoi{10.1088/0004-637X/798/2/105}

\bibitem[{{Basu} {et~al.}(2019{\natexlab{b}}){Basu}, {Paul}, \&
  {Mitra}}]{BPM19}
{Basu}, R., {Paul}, A., \& {Mitra}, D. 2019{\natexlab{b}}, \mnras, 486, 5216,
  \dodoi{10.1093/mnras/stz1225}

\bibitem[{{Champion} {et~al.}(2005){Champion}, {Lorimer}, {McLaughlin},
  {Xilouris}, {Arzoumanian}, {Freire}, {Lommen}, {Cordes}, \& {Camilo}}]{CLM05}
{Champion}, D.~J., {Lorimer}, D.~R., {McLaughlin}, M.~A., {et~al.} 2005,
  \mnras, 363, 929, \dodoi{10.1111/j.1365-2966.2005.09499.x}

\bibitem[{{Geppert}(2017)}]{G17}
{Geppert}, U. 2017, Journal of Astrophysics and Astronomy, 38, 46,
  \dodoi{10.1007/s12036-017-9460-y}

\bibitem[{{Gil} {et~al.}(1984){Gil}, {Gronkowski}, \& {Rudnicki}}]{GGR84}
{Gil}, J., {Gronkowski}, P., \& {Rudnicki}, W. 1984, \aap, 132, 312

\bibitem[{{Gil} {et~al.}(2008){Gil}, {Haberl}, {Melikidze}, {Geppert}, {Zhang},
  \& {Melikidze}}]{GHM08}
{Gil}, J., {Haberl}, F., {Melikidze}, G., {et~al.} 2008, \apj, 686, 497,
  \dodoi{10.1086/590657}

\bibitem[{{Gil} {et~al.}(2003){Gil}, {Melikidze}, \& {Geppert}}]{GMG03}
{Gil}, J., {Melikidze}, G.~I., \& {Geppert}, U. 2003, \aap, 407, 315,
  \dodoi{10.1051/0004-6361:20030854}

\bibitem[{{Gil} {et~al.}(2002){Gil}, {Melikidze}, \& {Mitra}}]{GMM02}
{Gil}, J.~A., {Melikidze}, G.~I., \& {Mitra}, D. 2002, \aap, 388, 235,
  \dodoi{10.1051/0004-6361:20020473}

\bibitem[{{Hermsen} {et~al.}(2013){Hermsen}, {Hessels}, {Kuiper}, {van
  Leeuwen}, {Mitra}, {de Plaa}, {Rankin}, {Stappers}, {Wright}, \& {et
  al.}}]{H13}
{Hermsen}, W., {Hessels}, J.~W.~T., {Kuiper}, L., {et~al.} 2013, Science, 339,
  436, \dodoi{10.1126/science.1230960}

\bibitem[{{Hermsen} {et~al.}(2018){Hermsen}, {Kuiper}, {Basu}, {Hessels},
  {Mitra}, {Rankin}, {Stappers}, {Wright}, {Grie{\ss}meier}, {Serylak},
  {Horneffer}, {Tiburzi}, \& {Ho}}]{HKB18}
{Hermsen}, W., {Kuiper}, L., {Basu}, R., {et~al.} 2018, \mnras, 480, 3655,
  \dodoi{10.1093/mnras/sty2075}

\bibitem[{{Kijak} \& {Gil}(1998)}]{KG98}
{Kijak}, J., \& {Gil}, J. 1998, \mnras, 299, 855,
  \dodoi{10.1046/j.1365-8711.1998.01832.x}

\bibitem[{{Krzeszowski} {et~al.}(2009){Krzeszowski}, {Mitra}, {Gupta}, {Kijak},
  {Gil}, \& {Acharyya}}]{KMG09}
{Krzeszowski}, K., {Mitra}, D., {Gupta}, Y., {et~al.} 2009, \mnras, 393, 1617,
  \dodoi{10.1111/j.1365-2966.2008.14287.x}

\bibitem[{{Mitra}(2017)}]{M17}
{Mitra}, D. 2017, Journal of Astrophysics and Astronomy, 38, 52,
  \dodoi{10.1007/s12036-017-9457-6}

\bibitem[{{Mitra} {et~al.}(2016){Mitra}, {Basu}, {Maciesiak}, {Skrzypczak},
  {Melikidze}, {Szary}, \& {Krzeszowski}}]{MBM16}
{Mitra}, D., {Basu}, R., {Maciesiak}, K., {et~al.} 2016, \apj, 833, 28,
  \dodoi{10.3847/1538-4357/833/1/28}

\bibitem[{{Mitra} {et~al.}(2020){Mitra}, {Basu}, {Melikidze}, \&
  {Arjunwadkar}}]{MBM20}
{Mitra}, D., {Basu}, R., {Melikidze}, G.~I., \& {Arjunwadkar}, M. 2020, \mnras,
  492, 2468, \dodoi{10.1093/mnras/stz3620}

\bibitem[{{Mitra} \& {Deshpande}(1999)}]{MD99}
{Mitra}, D., \& {Deshpande}, A.~A. 1999, \aap, 346, 906.
\newblock \doarXiv{astro-ph/9904336}

\bibitem[{{Mitra} \& {Li}(2004)}]{ML04}
{Mitra}, D., \& {Li}, X.~H. 2004, \aap, 421, 215,
  \dodoi{10.1051/0004-6361:20034094}

\bibitem[{{Mitra} \& {Rankin}(2002)}]{ET_MR02}
{Mitra}, D., \& {Rankin}, J.~M. 2002, \apj, 577, 322, \dodoi{10.1086/342136}

\bibitem[{{P{\'e}tri} \& {Mitra}(2020)}]{PM20}
{P{\'e}tri}, J., \& {Mitra}, D. 2020, \mnras, 491, 80,
  \dodoi{10.1093/mnras/stz2974}

\bibitem[{{Radhakrishnan} \& {Cooke}(1969)}]{RC69}
{Radhakrishnan}, V., \& {Cooke}, D.~J. 1969, \aplett, 3, 225

\bibitem[{{Rankin}(1993)}]{ET_R93}
{Rankin}, J.~M. 1993, \apj, 405, 285, \dodoi{10.1086/172361}

\bibitem[{{Ruderman} \& {Sutherland}(1975)}]{RS75}
{Ruderman}, M.~A., \& {Sutherland}, P.~G. 1975, \apj, 196, 51,
  \dodoi{10.1086/153393}

\bibitem[{{Shang} {et~al.}(2022){Shang}, {Bai}, {Dang}, \& {Zhi}}]{SBD22}
{Shang}, L.-H., {Bai}, J.-T., {Dang}, S.-J., \& {Zhi}, Q.-J. 2022, Research in
  Astronomy and Astrophysics, 22, 025018, \dodoi{10.1088/1674-4527/ac424d}

\bibitem[{{Skrzypczak} {et~al.}(2018){Skrzypczak}, {Basu}, {Mitra},
  {Melikidze}, {Maciesiak}, {Koralewska}, \& {Filothodoros}}]{SBM18}
{Skrzypczak}, A., {Basu}, R., {Mitra}, D., {et~al.} 2018, \apj, 854, 162,
  \dodoi{10.3847/1538-4357/aaa758}

\bibitem[{{Song} {et~al.}(2023){Song}, {Weltevrede}, {Szary}, {Wright},
  {Keith}, {Basu}, {Johnston}, {Karastergiou}, {Main}, {Oswald},
  {Parthasarathy}, {Posselt}, {Bailes}, {Buchner}, {Hugo}, \&
  {Serylak}}]{SWS23}
{Song}, X., {Weltevrede}, P., {Szary}, A., {et~al.} 2023, \mnras, 520, 4562,
  \dodoi{10.1093/mnras/stad135}

\bibitem[{{Szary} {et~al.}(2017){Szary}, {Gil}, {Zhang}, {Haberl}, {Melikidze},
  {Geppert}, {Mitra}, \& {Xu}}]{SGZ17}
{Szary}, A., {Gil}, J., {Zhang}, B., {et~al.} 2017, \apj, 835, 178,
  \dodoi{10.3847/1538-4357/835/2/178}

\bibitem[{{Szary} \& {van Leeuwen}(2017)}]{SvL17}
{Szary}, A., \& {van Leeuwen}, J. 2017, \apj, 845, 95,
  \dodoi{10.3847/1538-4357/aa803a}

\bibitem[{{Szary} {et~al.}(2020){Szary}, {van Leeuwen}, {Weltevrede}, \&
  {Maan}}]{SvLW20}
{Szary}, A., {van Leeuwen}, J., {Weltevrede}, P., \& {Maan}, Y. 2020, \apj,
  896, 168, \dodoi{10.3847/1538-4357/ab9226}

\bibitem[{{Sznajder} \& {Geppert}(2020)}]{SG20}
{Sznajder}, M., \& {Geppert}, U. 2020, \mnras, 493, 3770,
  \dodoi{10.1093/mnras/staa492}

\bibitem[{{von Hoensbroech} \& {Xilouris}(1997)}]{vHX97}
{von Hoensbroech}, A., \& {Xilouris}, K.~M. 1997, \aaps, 126, 121

\bibitem[{{Weltevrede}(2016)}]{W16}
{Weltevrede}, P. 2016, \aap, 590, A109, \dodoi{10.1051/0004-6361/201527950}

\bibitem[{{Weltevrede} {et~al.}(2006){Weltevrede}, {Edwards}, \&
  {Stappers}}]{WES06}
{Weltevrede}, P., {Edwards}, R.~T., \& {Stappers}, B.~W. 2006, \aap, 445, 243,
  \dodoi{10.1051/0004-6361:20053088}

\bibitem[{{Weltevrede} \& {Johnston}(2008)}]{WJ08}
{Weltevrede}, P., \& {Johnston}, S. 2008, \mnras, 391, 1210,
  \dodoi{10.1111/j.1365-2966.2008.13950.x}

\end{thebibliography}
\bibliographystyle{aasjournal}

\appendix

\section{Estimating the Nature of Polar Cap and the Sparking Distribution} \label{app:psgsprk}

\begin{figure}
\gridline{\fig{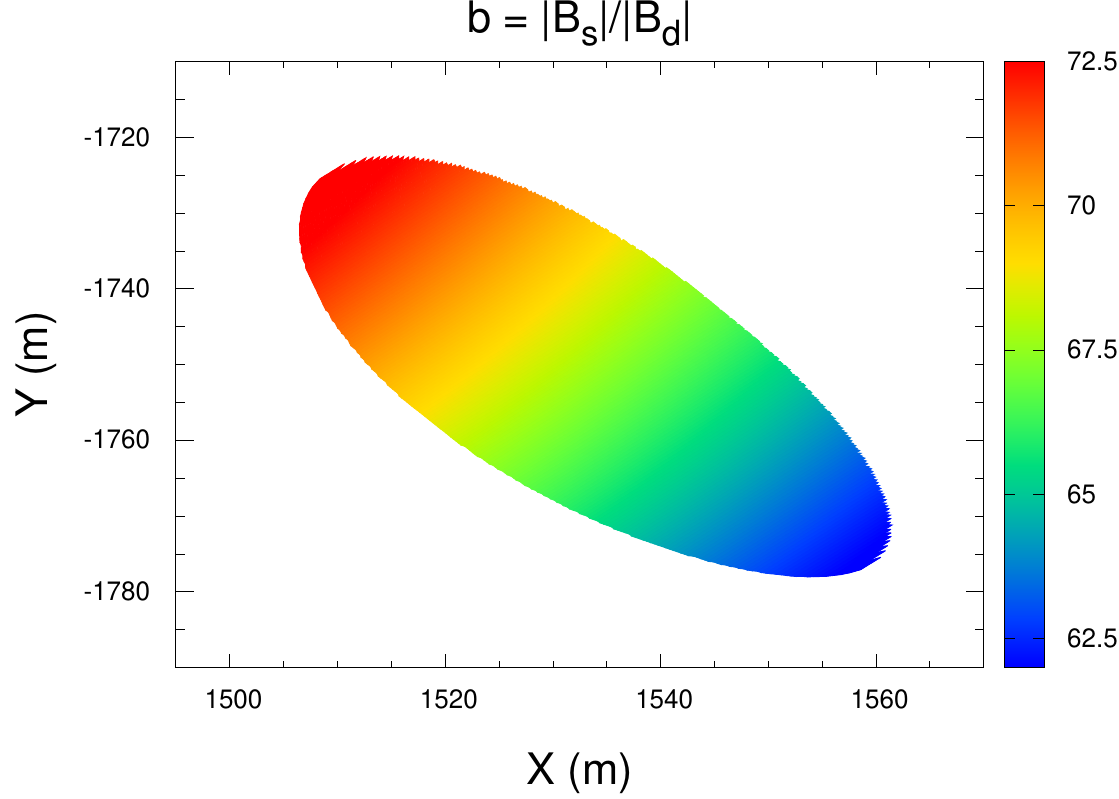}{0.45\textwidth}{(a)}
          \fig{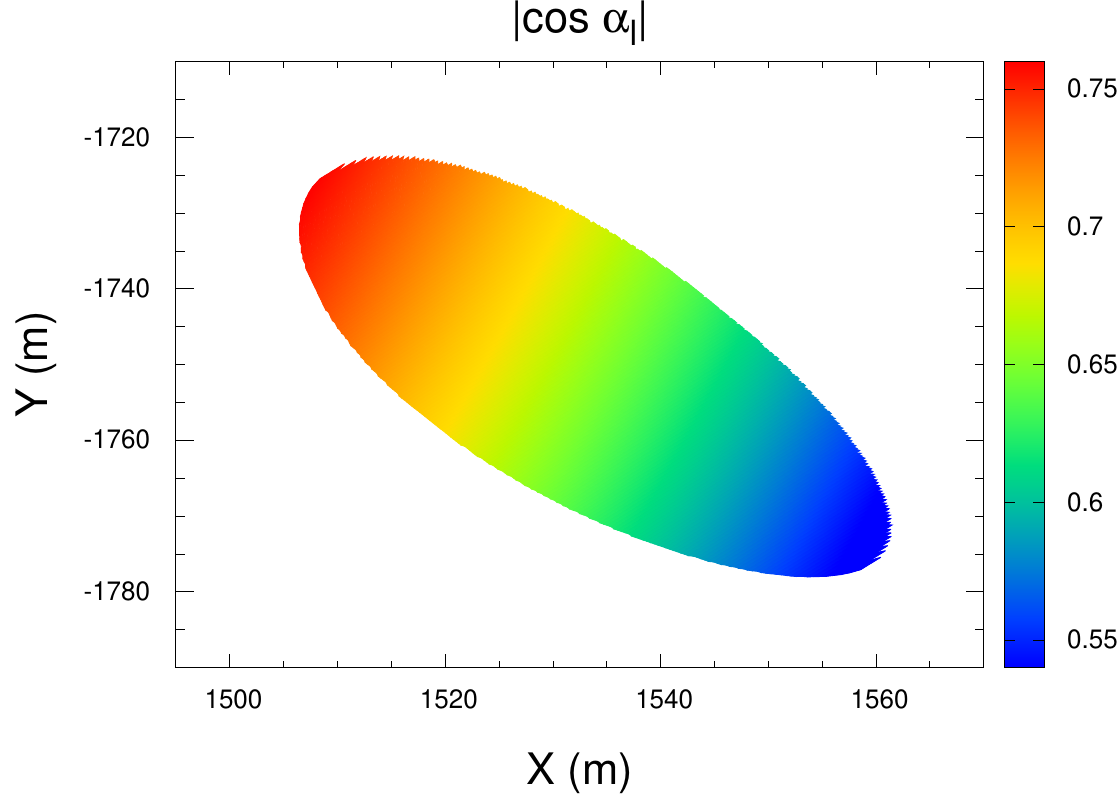}{0.45\textwidth}{(b)}}
\gridline{\fig{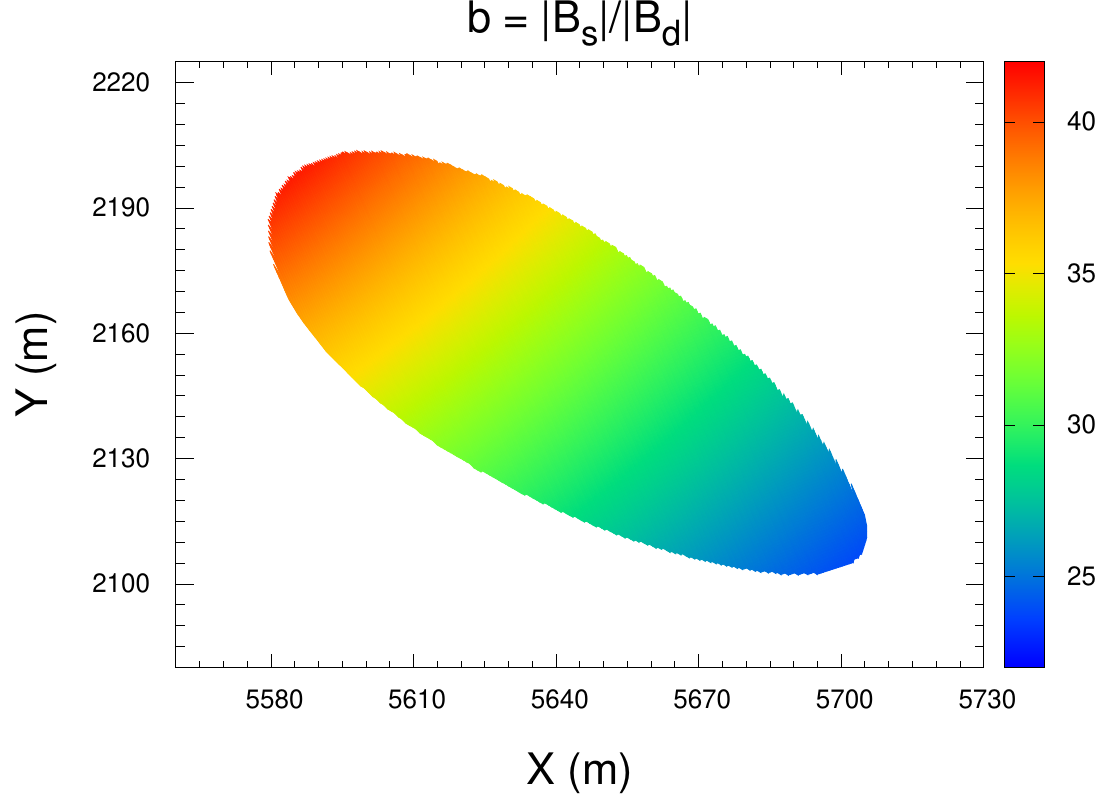}{0.45\textwidth}{(c)}
          \fig{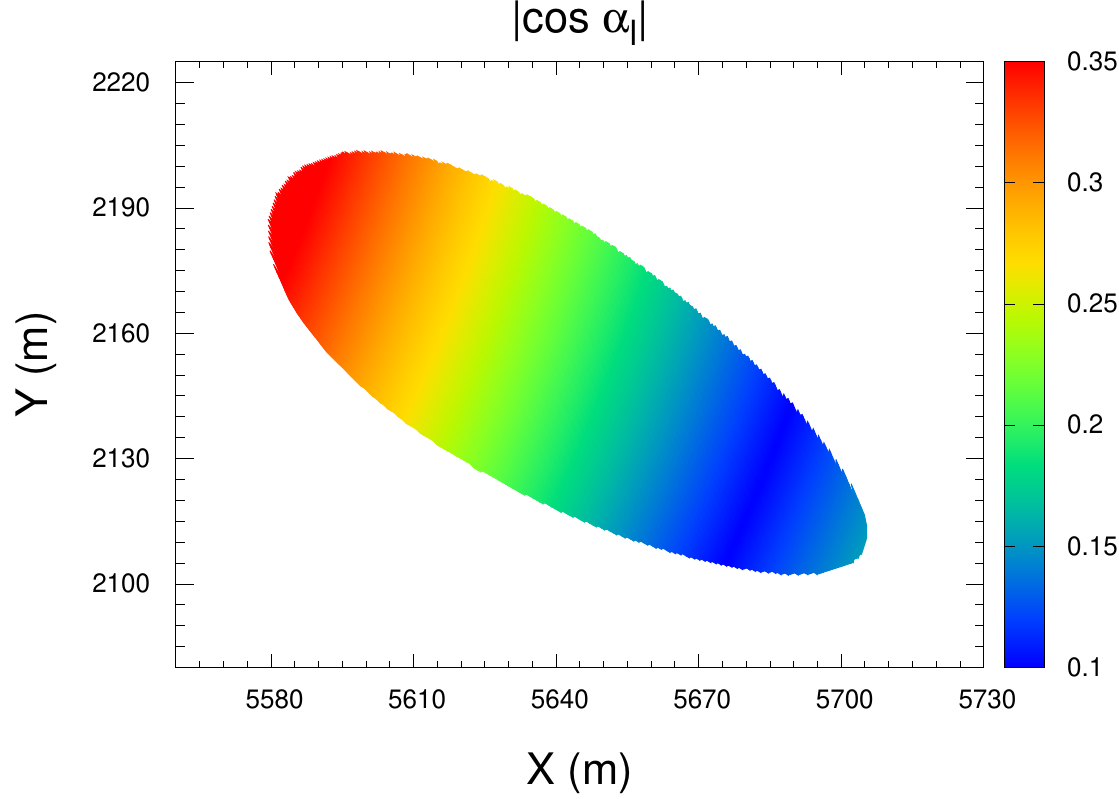}{0.45\textwidth}{(d)}}
\caption{The figure shows the physical conditions above the simulated polar 
caps of the two pulsars. 
\label{fig:polcap}}
\end{figure}

Fig.\ref{fig:polcap} shows the variations of the magnetic fields across the 
polar cap surface. Fig.\ref{fig:polcap}(a) and Fig.\ref{fig:polcap}(c) 
corresponds to the parameter $b$ which specify the relative strength of the 
surface field while Fig.\ref{fig:polcap}(b) and Fig.\ref{fig:polcap}(d) shows
$|\cos{\alpha_l}|$ which shows the local inclination of the magnetic field. 
Both these quantities determine the screening factor, the surface temperature
and the potential difference in PSG estimated in Table \ref{tab:capfit}. 

\begin{figure}
\begin{interactive}{animation}{Bidrift_spark.mp4}
\epsscale{0.55}
\plotone{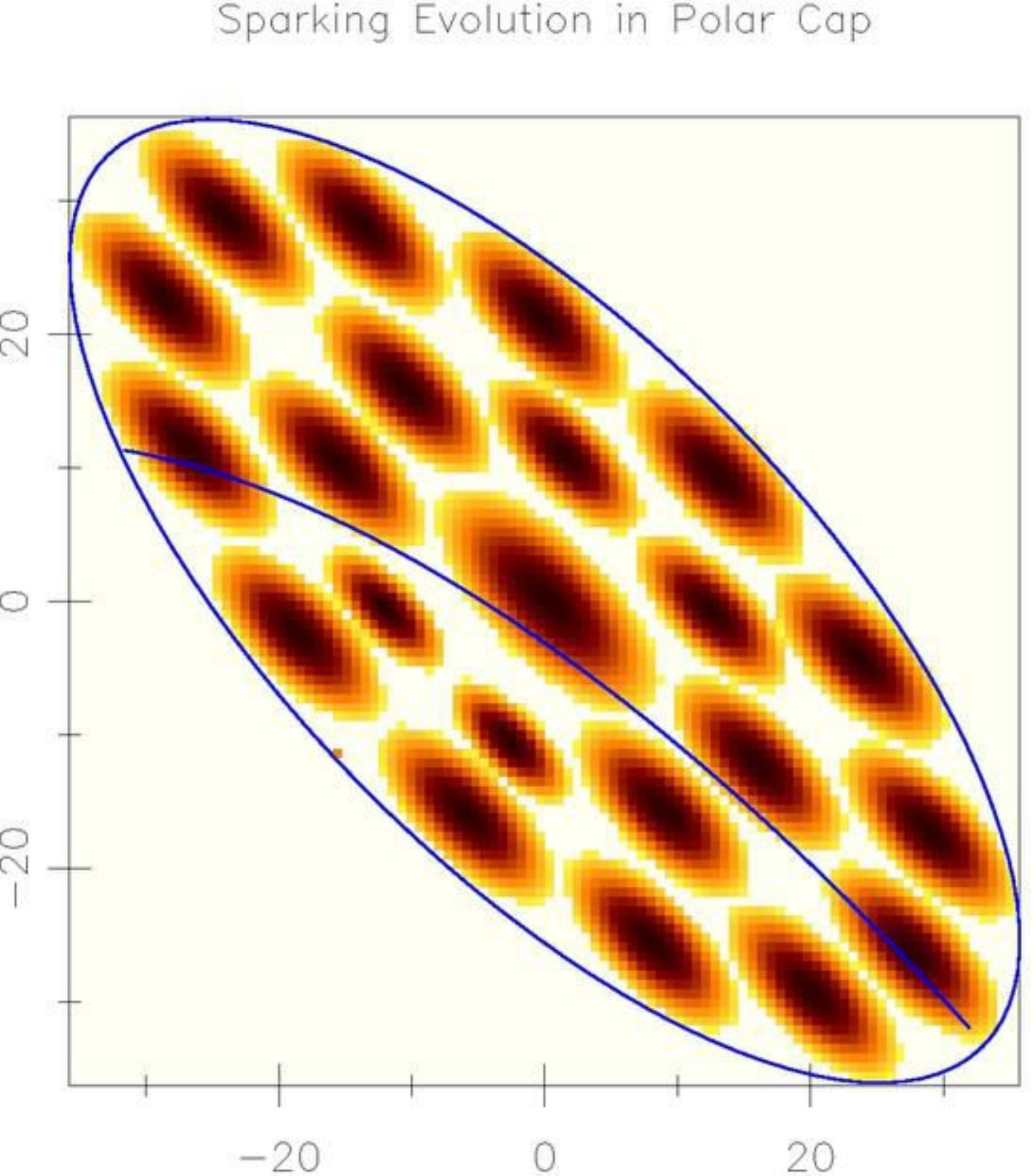}
\end{interactive}
\caption{The figure shows the two dimensional distribution of the sparking
pattern above the simulated polar cap corresponding to PSR J1034$-$3224. The
sparks are arranged in two concentric elliptical annulus around a central spark
in a tightly packed configuration. The sparking pattern evolves with time to
show a anti-clockwise shift in the left half and an clockwise shift in the 
right half bounded by the points $\theta'_s = 88.7\degr$ and $\theta'_e = 
268.6\degr$, where the pattern shifts away from $\theta'_s$ and converges
towards $\theta'_e$. The line of sight (LOS) traverse across the emission beam
at a relative shift $\beta=1.6\degr$ from the center and its imprint on the 
polar cap is also shown. The dynamical evolution of the spark distribution 
across the LOS results in bi-drifting. \\\\ An animation showing the evolution 
of the spark configuration with time is available.
\label{fig:sparkbidrift}}
\end{figure}

\begin{figure}
\begin{interactive}{animation}{coh_spark.mp4}
\epsscale{0.55}
\plotone{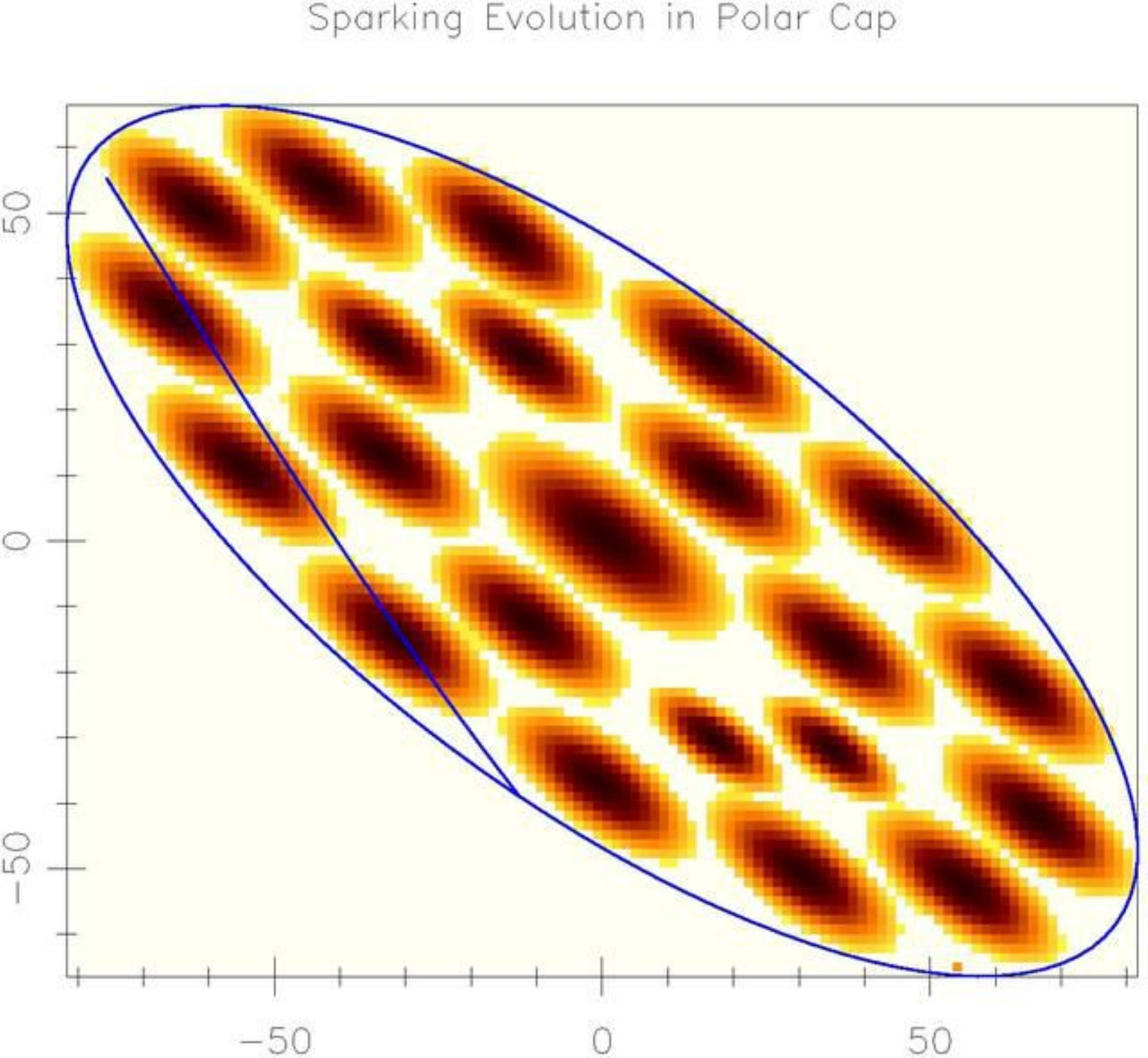}
\end{interactive}
\caption{The figure shows the two dimensional distribution of the sparking
pattern above the simulated polar cap corresponding to PSR J1720$-$2933. The
sparks are arranged in two concentric elliptical annulus around a central spark
in a tightly packed configuration. The sparking pattern evolves with time to
show a anti-clockwise shift in the left half and an clockwise shift in the
right half bounded by the points $\theta'_s = 122.4\degr$ and $\theta'_e = 
302.4\degr$, where the pattern shifts away from $\theta'_s$ and converges
towards $\theta'_e$. The line of sight (LOS) traverse across the emission beam
at a relative shift $\beta=-5.4\degr$ from the center and its imprint on the
polar cap is also shown. The dynamical evolution of the spark distribution
across the LOS results in coherent subpulse drifting. \\\\ An animation showing
the evolution of the spark configuration with time is available.
\label{fig:sparkcoh}}
\end{figure}

The two dimensional distribution of the sparks and their evolution with time 
in PSR J1034$-$3224 is shown in animated Fig.\ref{fig:sparkbidrift}, along with
the LOS cut across the cap which results in the observed bi-drifting behaviour.
The animated Fig.\ref{fig:sparkcoh} shows the spark distribution in polar cap 
of PSR J1720$-$2933, along with the LOS geometry that gives rise to the 
coherent drifting behaviour. In each case we have considered two concentric
rings for the spark evolution around a centrally localized spark, which evolve 
in two separate directions in the clockwise and counter-clockwise manner 
respectively. The bi-drifting behaviour requires the LOS to cut across both 
these tracks in both directions of evolution in an appropriate manner. The 
coherent drifting on the other requires the LOS to only follow the outer track 
along one direction.

\begin{deluxetable}{cccccccccccc}
\tablecaption{The details of the spark distribution in polar cap \label{tab:sparkdist}}
\tablewidth{0pt}
\tablehead{
    & \colhead{Cone}  & \colhead{$i$} & \colhead{$a_{out}$} & \colhead{$a_{in}$} & \colhead{$b_{out}$} & \colhead{$b_{in}$} & \colhead{$N_{sprk}$} & \colhead{$\theta_{sprk}$} & \colhead{$a_{trk}$} & \colhead{$b_{trk}$} & \colhead{$\omega_{u,d}$}  \\
   &   &   & \colhead{(m)} & \colhead{(m)} & \colhead{(m)} & \colhead{(m)} &   & \colhead{($\degr$)} & \colhead{(m)} & \colhead{(m)} & \colhead{(deg s$^{-1}$)}
          }
\startdata
   &  &  &  &  &  &  &  &  &  &  &  \\
 J1034$-$3224 & Outer & 1 & 36.2 & 22.8 & 15.1 & 9.5 & 13 & 27.7 & 29.5 & 12.3 & $\mp3.8$ \\
   & Inner & 2 & 22.8 & 9.4 & 9.5 & 3.9 & 7 & 51.4 & 16.1 & 6.7 & $\mp7.1$ \\
   &  &  &  &  &  &  &  &  &  &  &  \\
 J1720$-$2933 & Outer & 1 & 75.2 & 47.3 & 30.1 & 19.0 & 13 & 27.7 & 61.3 & 24.5 & $\mp11.3$ \\
   & Inner & 2 & 47.3 & 19.5 & 30.1 & 7.8 & 7 & 51.4 & 33.4 & 13.4 & $\mp21.0$ \\
   &  &  &  &  &  &  &  &  &  &  &  \\
\enddata
\end{deluxetable}

Table \ref{tab:sparkdist} reports the different parameters that describe the 
the distribution of sparks in each polar cap \citep[see][for a more detailed 
description]{BMM22}. The sparks are tightly packed in the IAR and arranged in 
concentric rings around a central spark. The observations of the average beam 
properties show the presence of two pairs of conal components surrounding a 
central core that constrains the typical number of such tracks to be $N_{trk} =
2$. The maximum packing condition specifies the size of the sparks to be 
$N_{trk} = {\rm Int}({a_{cap}/a_{sprk}}) = {\rm Int}({b_{cap}/c_{sprk}})$. In 
an elliptical polar cap the tracks are bound by the ellipses defined by 
$a_{out}^i$, $b_{out}^i$ and $a_{in}^i$, $b_{in}^i$ where $i = 1,2$, 
corresponding to the outer and inner conal tracks respectively. The maximum 
number of fully formed spark along each conal track is $N_{sprk}^i$ and these 
quantities can be estimated as : 
\begin{eqnarray}
a_{out}^i = a_{cap} - 2(i-1) a_{sprk},~a_{in}^i = a_{out}^i - 2a_{sprk}, \nonumber\\
b_{out}^i = b_{cap} - 2(i-1) b_{sprk},~b_{in}^i = b_{out}^i - 2b_{sprk}, \nonumber\\ 
N_{sprk}^i = {\rm Int}\left(F (a_{out}^i b_{out}^i - a_{in}^i b_{in}^i)/(a_{sprk} b_{sprk}))\right). 
\end{eqnarray}
Here $F$ is a scaling factor for maximum packing which was find to be around 
0.75. The angular size of the sparks in each track is $\theta_{sprk}^i = 
2\pi/N_{sprk}^i$, and their centers lie on the ellipse specified by $a_{trk}^i 
= (a_{out}^i + a_{in}^i)/2$ and $b_{trk}^i = (b_{out}^i + b_{in}^i)/2$. 

In order to estimate the two dimensional spark distribution a Cartesian 
$x'y'$-plane is defined to contain the elliptical polar cap with origin at the 
center of the ellipse. The boundary of the upper and lower halves of the polar 
cap signifying clockwise and anti-clockwise evolution of the sparking pattern 
is specified by the angles $\theta'_s = \pi/2 - \phi_{cap}^c$ and $\theta'_e = 
3\pi/2 - \phi_{cap}^c$, here $\theta'$ is the polar angle in the $x'y'$-plane. 
The sparking pattern diverges away from $\theta'_s$ and converges toward
$\theta'_e$ (see Fig.\ref{fig:sparkbidrift} and \ref{fig:sparkcoh}). The rate 
of shifting of the patterns in the two halves of each ring is estimated as 
$\omega_{u,d}$ = $\mp\theta_{sprk}/P_3$.

\section{Simulated Single Pulse sequence with Subpulse Drifting} \label{app:sinlrfs}

\begin{figure}
\gridline{\fig{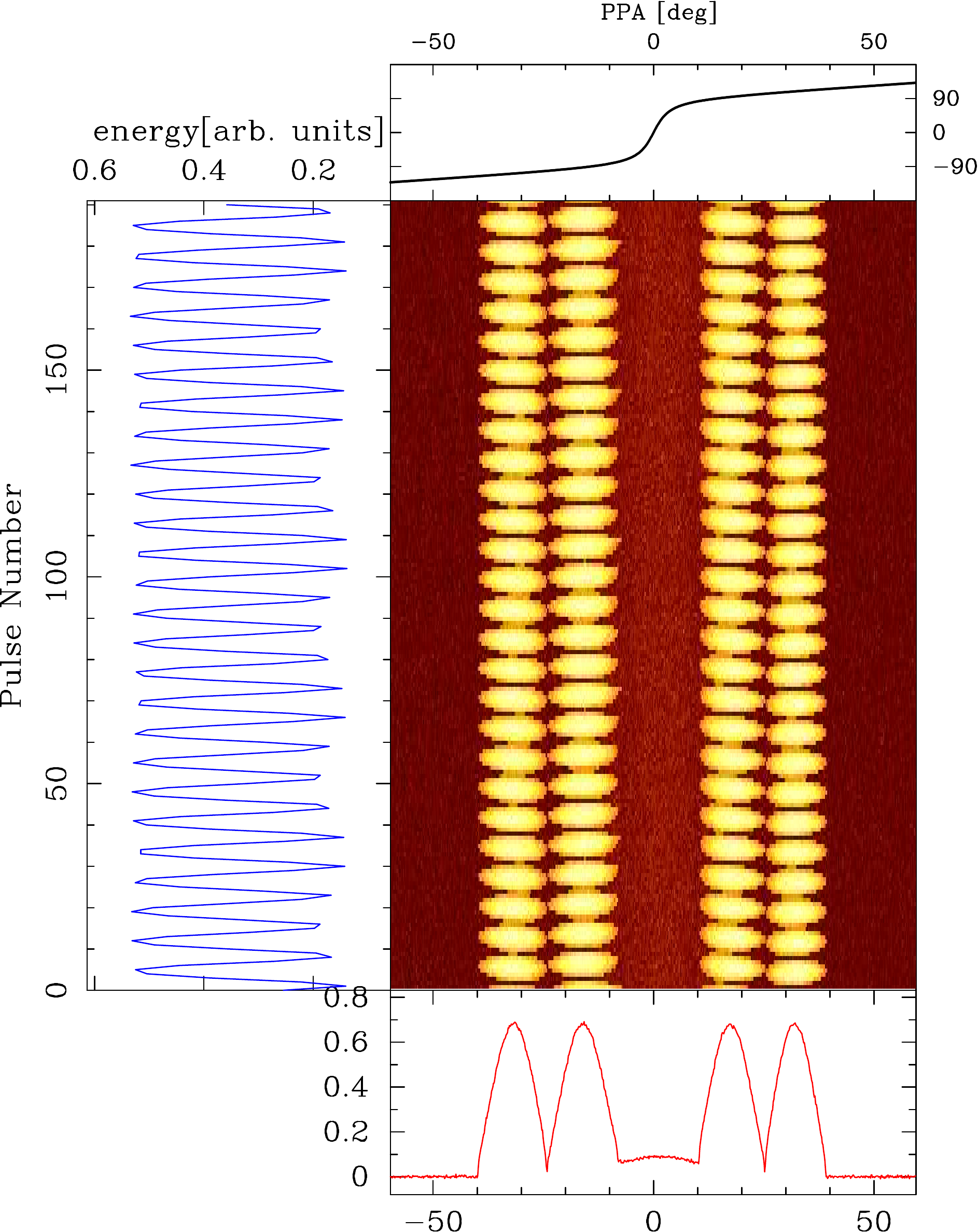}{0.42\textwidth}{(a)}
          \fig{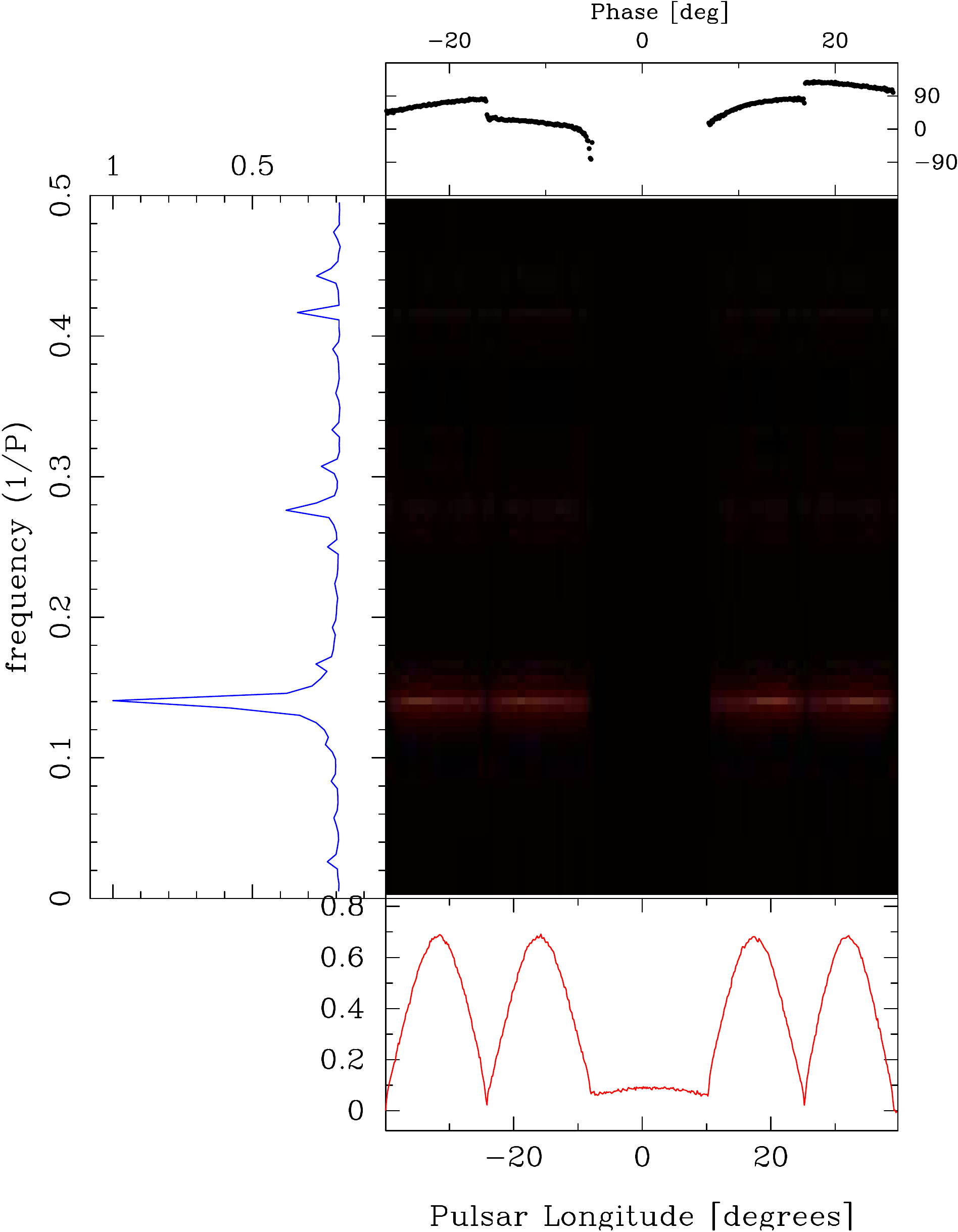}{0.42\textwidth}{(b)}}
\gridline{\fig{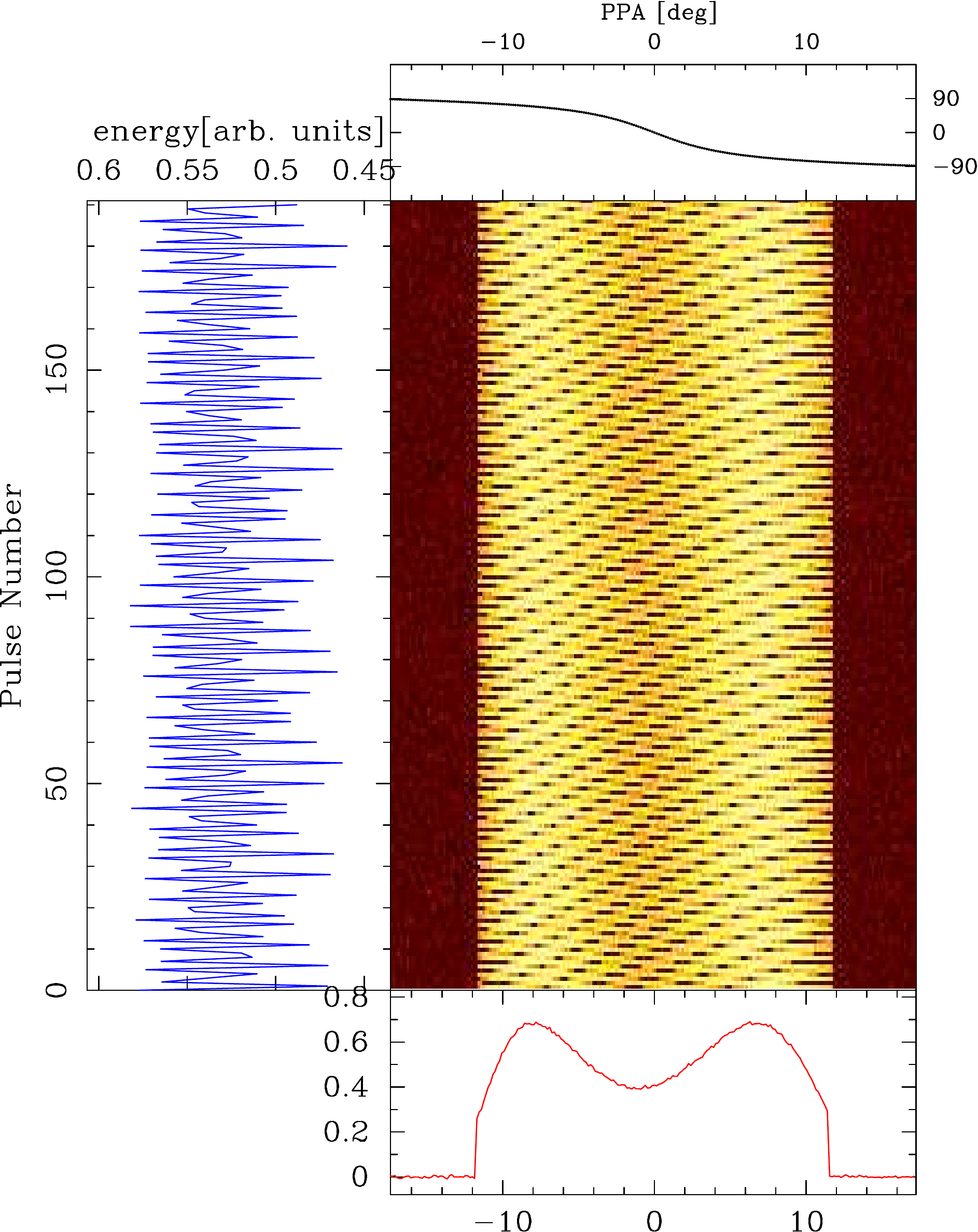}{0.42\textwidth}{(c)}
          \fig{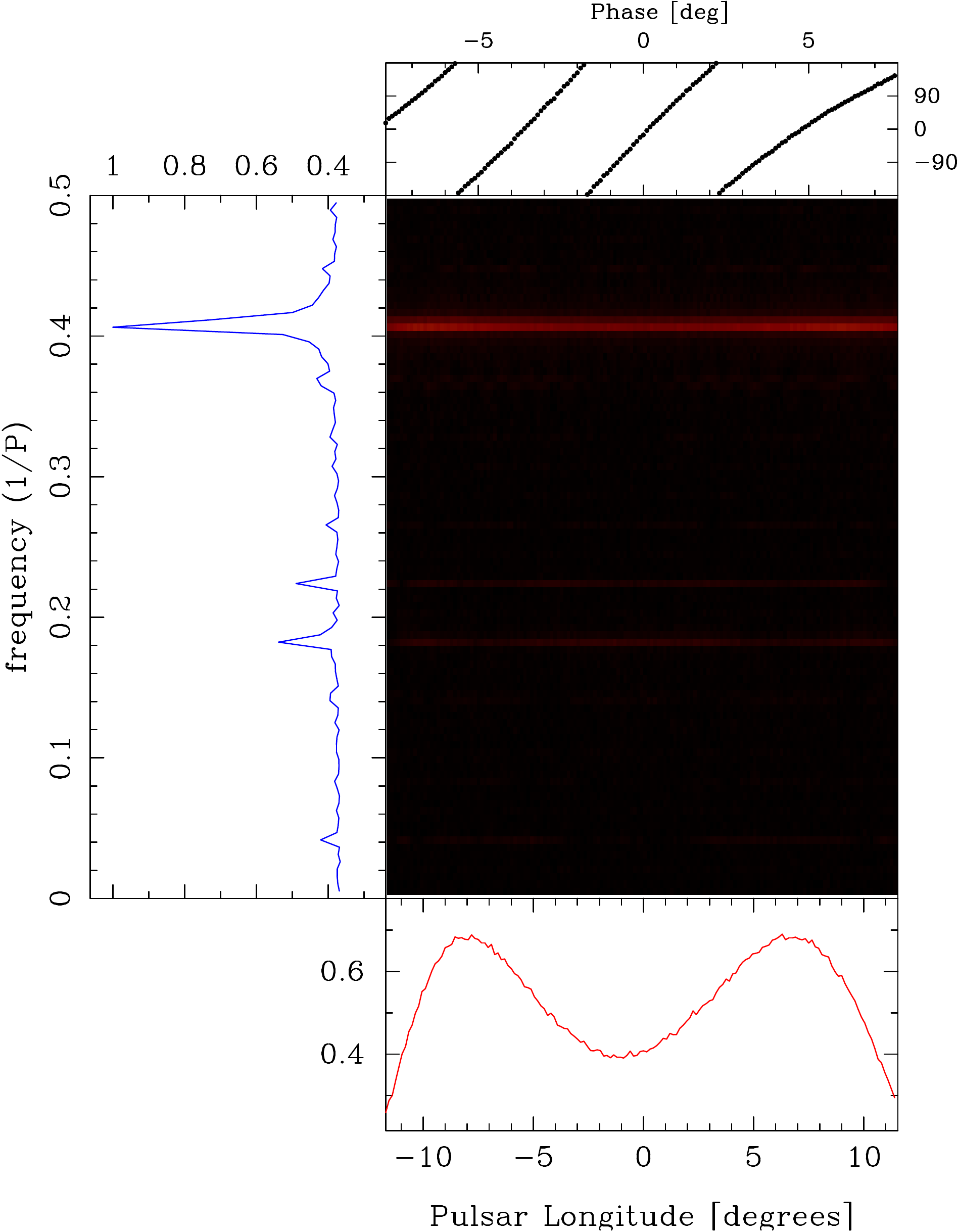}{0.42\textwidth}{(d)}}
\caption{The figure shows single pulse simulations demonstrating the 
bi-drifting behaviour of PSR J1034$-$3224~in the two upper panels (a), (b) and 
coherent drifting behaviour of PSR J1720$-$2933~in the lower panels (c), (d). 
Pulse stack with 200 simulated single pulses are shown in (a) and (c), while 
the Longitude Resolved Fluctuation Spectra (LRFS) across the pulse window is 
represented in (b) and (d). The drifting periodicity, $P_3$ = 7.2$P$ is seen as
peak frequency, $f_p\sim0.14$ cycles/$P$ in panel (b) while $P_3$ = 2.45$P$ is 
seen as peak frequency, $f_p\sim0.41$ cycles/$P$ in panel (d). The evolution of
the sparking pattern across the polar cap is reflected in the phase behaviour 
across the profile in the right panels.
\label{fig:singlrfs}}
\end{figure}

The procedure for simulating the single pulse emission as well as estimating 
the LRFS to obtain the subpulse drifting behaviour in the two pulsars is 
described in \citet{BMM22}. Fig.\ref{fig:singlrfs}(a) shows around 200 pulses
resembling the bi-drifting behaviour in PSR J1034$-$3224 while 
Fig.\ref{fig:singlrfs}(b) shows the estimated LRFS on this sequence reflecting 
the drifting peak frequency and the phase variations across the profile. The 
LOS grazes near the edge of the central spark and as a result the central 
component is not visible in the average profile which comprises of four 
components, matching the observations. Fig.\ref{fig:singlrfs}(c) shows around 
200 pulses resembling the coherent drifting behaviour in PSR J1720$-$2933. The 
LRFS in Fig.\ref{fig:singlrfs}(d) shows the frequency peak corresponding to 
periodic drifting and the large phase behaviour across the emission window. The
LOS only cuts along the outer conal track of the sparks and the average profile
shows a dip at the center which is the observed behaviour.

\end{document}